\documentclass[useAMS,usenatbib]{mn2e}
\usepackage{graphicx}
\def\apj{ApJ}
\def\apjl{ApJL}
\def\mnras{MNRAS}
\def\aap{AAP}

\def\swift{{\it Swift~}}
\def\fermi{{\it Fermi~}}
\def\suzaku{{\it Suzaku~}}

\title[Fundamental plane of GRBs and Cosmology]
{Cosmological Constraints from calibrated Yonetoku and Amati 
relation suggest Fundamental plane of Gamma-ray bursts}
\author[R. Tsutsui et al.]
{Ryo Tsutsui$^{1}$\thanks{E-mail: tsutsui@tap.scphys.kyoto-u.ac.jp (RT)},
Takashi Nakamura$^{1}$, Daisuke Yonetoku$^{2}$,Toshio Murakami$^{2}$,
\newauthor Yoshiki Kodama$^{2}$,
and Keitaro Takahashi$^{3}$\\
$^{1}$Department of Physics, Kyoto University,
Kyoto 606-8502, Japan\\
$^{2}$Department of Physics, Faculty of Science,
Kanazawa University, Kakuma, Kanazawa, Ishikawa 920-1192, Japan\\
$^{3}$Yukawa Institute for Theoretical Physics, Kyoto University, 
Kyoto 606-8502, Japan}

\begin{document}


\pagerange{\pageref{firstpage}--\pageref{lastpage}} \pubyear{2008}

\maketitle

\label{firstpage}

\begin{abstract}
We consider two empirical relations using data only from
the prompt emission of Gamma-Ray Bursts (GRBs),
the peak energy ($E_p$) - peak luminosity ($L_p$) relation
(so called Yonetoku relation) and the $E_p$-isotropic energy
($E_{\rm iso}$) relation (so called Amati relation).
Both relations show high correlation degree, 
but they also have larger dispersion around the best fit function 
rather than the statistical uncertainty.
Then we first investigated the correlation between the residuals of 
$L_p$ and $E_{iso}$ from the best function, and found that 
a partial linear correlation degree is quite small of 
$\rho_{L_{p}~E_{iso} \cdot E_p}=0.379$. This fact indicates that 
some kinds of independence may exist between Amati and 
Yonetoku relation even if they are characterized by 
the same physical quantity $E_p$, 
and similar quantities $L_p$ and $E_{iso}$ which mean the brightness 
of the prompt emission. Therefore we may have to recognize 
two relations as the independent distance indicators.
From this point of view, we compare constraints on cosmological
parameters, $\Omega_m$ and $\Omega_{\Lambda}$, using 
the Yonetoku and the Amati relation calibrated by low-redshift
GRBs with $z < 1.8$. We found that they are different
in 1-$\sigma$ level, although they are still consistent
in 2-$\sigma$ level. 
In this paper, we introduce a luminosity time $T_L$
defined by $T_L \equiv E_{\rm iso}/L_p$ as a hidden parameter
to correct the large dispersion of the Yonetoku relation. 
A new relation is described as
$(L_p/{10^{52}~\rm{erg~s^{-1}}}) =
 10^{-3.87\pm0.19}(E_p/{\rm{keV}})^{1.82\pm0.08}
 (T_L/{\rm{s}})^{-0.34\pm0.09}$. We succeeded in reducing 
the systematic error about 40\% level, 
and might be regarded as "Fundamental plane" of GRBs. 
We show a possible radiation 
model for this new relation. Finally, applying the new
relation to high-redshift GRBs with $1.8 < z < 5.6$,
we obtain $(\Omega_m,\Omega_{\Lambda}) =
 (0.17^{+0.15}_{-0.08},1.21^{+0.07}_{-0.61})$,
which is consistent with the concordance cosmological model.

\end{abstract}


\section{Introduction}
\label{sec:intro}
In our previous papers \citep{Kodama2008,Tsutsui2008b},
we calibrated the relation between peak energy $E_p$
and peak luminosity $L_p$ of prompt GRB emission
(so called Yonetoku relation \citep{Yonetoku2004})
by 33 low-redshift GRBs ($z < 1.62$) whose luminosity distances
were estimated from SNeIa \citep{Riess2007,Wood-Vasey2007,Davis2007}. 
Then we used the calibrated Yonetoku relation as a distance indicator
like the period-luminosity relation of Cepheid variables,
and extended the Hubble diagram up to $z = 5.6$.
\citet{Tsutsui2008b} showed that GRBs constrain cosmological
parameters in a different way from SNeIa, and GRBs could be
useful to probe cosmological expansion of high-redshift
universe where no SNIa has been observed.

The Amati relation \citep{Amati2002} is another relation
for prompt emission property. It involves peak energy $E_p$ and
isotropic energy $E_{\rm{iso}}$, and was originally derived
under a given set of cosmological parameters. Therefore,
the circularity problem arises if one applies
naively the Amati relation to determine cosmological parameters.
To overcome this difficulty, in this paper, we first calibrate
the Amati relation as we did for Yonetoku relation without 
assuming any cosmological models but using luminosity distance 
given by SNeIa for $z < 1.8$.

Although the strong correlation between $E_{\rm iso}$ and $L_p$ 
are generally confirmed, we found little trend between 
their data residuals of $\Delta L_p$ and $\Delta E_{\rm iso}$ 
from the best fit function of each relation.
We suggest statistical independence of Amati and Yonetoku 
relations while these two relations have the same parameter 
$E_p$ (\S-2).

Next, we will extend the Hubble diagram with the obtained
relations and make constraints on density parameters,
($\Omega_m,\Omega_{\Lambda}$). It is shown that the two
Hubble diagrams differ systematically at high redshifts and,
as a result, two different constraints are obtained (\S-3).
Although the difference is not so significant (1-$\sigma$ level),
taking the relatively large systematic errors in the two relations
themselves into consideration, it is suggested that there
may be a hidden parameter which characterizes the prompt
emission and reduces the systematic error of the distance indicator.
We introduce the luminosity time ($T_L \equiv E_{\rm iso}/L_p$)
as a possible hidden parameter to derive the $E_p$-$T_L$-$L_p$ relation, and
put constraints on cosmological parameters (\S-4).
Finally we give some comments on other relations used as
distance indicators in the past, and argue the advantage of
our new relation (\S-5). Throughout the paper, we fix
the current Hubble parameter as $H_0 = 66~{\rm km~s^{-1}Mpc^{-1}}$.

We note that, in whole  this paper,  tentatively we do not take possible selection effects and evolution effects 
on relations into accounts although a hot  debate 
 \citep{Butler2007,Li2007, Basilakos2008} exists. We think that the final resolution needs
the increase of the events with enough well determined parameters such as 
redshifts, $E_p$, $L_p$ and $E_{\rm{iso
}}$ since we should divide all data into several sub groups 
to check  possible selection effects and evolution effects.

\section{Calibration of Amati relation at low redshifts}

The typical spectrum of the prompt emission of GRBs can be
expressed as exponentially connected broken power-law,
so called Band function \citep{Band1993}. Then we can determine spectral peak
energy $E_p$, corresponding to the photon energy at maximum
in $\nu F_{\nu}$ spectra. There are two empirical relations
that relate prompt emission property with $E_p$. $E_p$-$E_{\rm iso}$
relation is the first one found by \citet{Amati2002}, which connects
$E_p$ with the isotropic equivalent energy $E_{\rm iso}$. The second one
is $E_p$-$L_p$ relation found by \citet{Yonetoku2004} which was used
in our previous papers \citep{Kodama2008,Tsutsui2008b}.

We first calibrate Amati relation in the same way as in our previous 
papers \citep{Kodama2008,Tsutsui2008b} and analyze the correlation of 
the residuals of GRB data from the relations and the partial correlation coefficient.
If there are a small degree of correlation between $L_p$ and $E_{iso}$ 
after removing the effect of $E_p$, it suggest the possible independence of the 
distance indicators, and might suggest the existence of a hidden 
parameter common to each relation. 

We found an empirical formula for the luminosity distance as a function
of redshift from 192 SNeIa observations
\citep{Riess2007,Davis2007,Wood-Vasey2007},
\begin{equation}
\frac{d_L}{10^{27}~\rm{cm}}
= 6.96 \times z^{1.79} + 14.79 \times z^{1.02}.
\label{eq:dL_SN}
\end{equation}
The reduced chi-square of the formula is $\chi_{\nu}^2 = 0.995$.
Note here that the formula is not unique and a different formula
is possible. Note also that we do not assume any cosmological
models at this stage, but simply assume that the Type Ia supernovae
are the standard candles for $0.168 < z < 1.755$. Furthermore,
we neglect the errors in Eq.~(\ref{eq:dL_SN}) in the following
analysis, which leads to the underestimation of errors
in cosmological parameters. Our purpose here is to
compare distance indicators and find a better indicator so that
this neglect would be reasonable.

\begin{figure}
\includegraphics[height=92mm,angle=270]{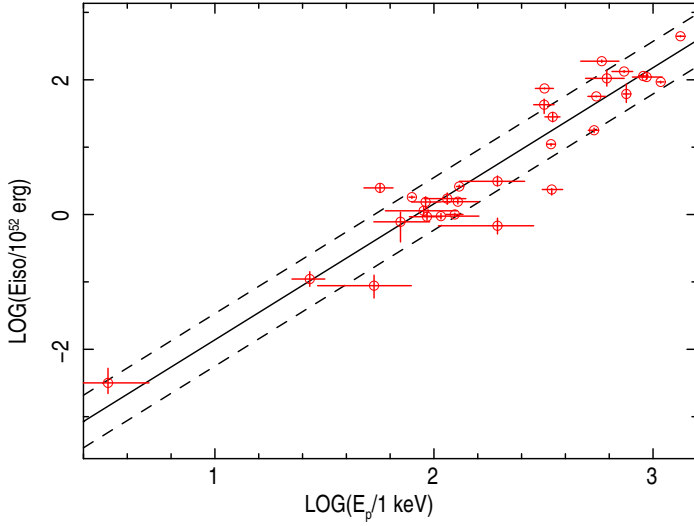}
\vspace{0pt}
\caption{The peak energy ($E_p$) and isotropic energy
($E_{\rm{iso}}$) of 31 GRBs with $z < 1.62$. The solid line
is the calibrated Amati relation given by Eq~(\ref{eq:Amati})
and dashed lines represent the 1-sigma region. The systematic error of 
 this relation is reduced to $\sigma_{sys}=0.35$ which is larger than 
 measurement errors.}
\label{fig1}
\end{figure}

We apply this formula to 31 low redshifts GRBs within $z < 1.6$.
( For details of our data, see \citet{Kodama2008}.)
In Fig.~\ref{fig1}  we show the peak energy $E_p$ and the isotropic
energy $E_{\rm{iso}}$ of 31 GRBs with $z < 1.62$. The solid line is
the calibrated Amati relation given by,
\begin{equation}
\frac{E_{\rm iso}}{10^{52}~\rm{erg}}
= 10^{-3.87 \pm 0.33}
  \left( \frac{E_p}{1\rm{keV}} \right)^{2.01 \pm 0.14},
\label{eq:Amati}
\end{equation}
where the statistical errors are indicated and
$E_{\rm iso} = 4 \pi d_{L}^2 S_{\rm bol}/(1+z)$ where $S_{\rm bol}$
is the bolometric fluence estimated in 1-10000~keV energy range
in GRB rest frame. The Pearson correlation coefficient is 0.943 and
the reduced chi-square is $\chi^2 = 28.5/29$ with the systematic error $\sigma_{sys}=0.35$.
Here we include not only errors in $E_{iso}$ but also errors in 
$E_p$, and the systematic error of this relation $\sigma_{sys}$, so the chi-square function is defined as
$
\chi^2(A,B)=\Sigma({\log{E_{iso}^{obs}}-A-B\log{E_p}})^2/(\sigma_{meas}^2+\sigma_{sys}^2)
$
where the weighting factor 
$\sigma^2_{meas}=\sigma_{\log{E_{iso}}}^2+B^2\sigma_{\log{E_p}}^2$.
The value of $\sigma_{sys}$ can be estimated by the value such that a 
$\chi^2$ fit to the Amati relation calibration produces a reduced 
$\chi^2$ of unity.

This kind of systematic error was also found in the Yonetoku
relation \citep{Kodama2008},
\begin{eqnarray}
\frac{L_p}{10^{52}~{\rm erg~s^{-1}}}
= 10^{-3.95 \pm 0.27}
  \left( \frac{E_p}{\rm 1 keV} \right)^{1.73 \pm 0.11},
\label{eq:Yonetoku}
\end{eqnarray}
where $L_p$ is 1-second peak luminosity. This Yonetoku relation
is slightly different from that in the previous work because we
include not only $L_p$ error but also $E_p$ error, and the systematic error in this paper.
The Pearson correlation coefficient is 0.948, the reduced chi-square is
$\chi^2=30.6/31$, and the systematic error is
$\sigma_{sys} = 0.27$.
The Amati relation has slightly larger systematic error
than the Yonetoku relation and, in both cases, systematic errors
are significantly larger than measurement errors.

\begin{figure}
\includegraphics[height=84mm,angle=270]{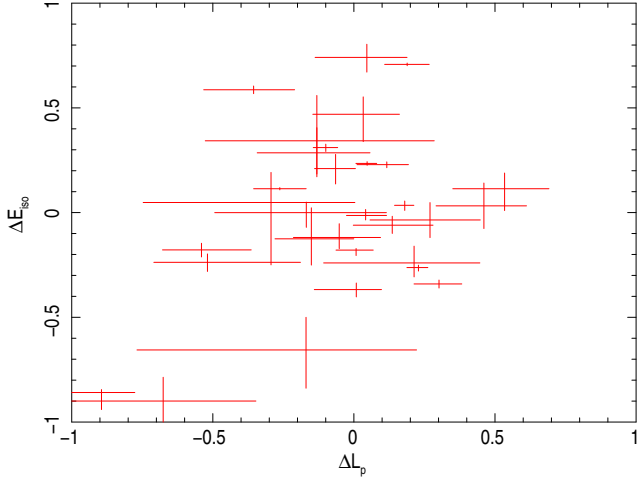}
\vspace{0pt}
\caption{Residuals from calibrated Yonetoku relation and Amati
relation from the observed $E_p$. Here, the residuals are
the differences between the observed quantities, $L_p^{\rm obs}$ and
$E^{\rm obs}_{\rm iso}$, and the expected quantities, $L_p^{\rm exp}$
and $E^{\rm exp}_{\rm iso}$, from Yonetoku and Amati relations
for the given observed $E_p$, respectively. There seems to be
no correlation, which implies the two relations are independent distance indicators.}
\label{fig2}
\end{figure}

Fig.~\ref{fig2} shows the correlation between two residuals 
which are defined as 
$\Delta L_{p} \equiv (\log L^{\rm obs}_{p} - \log L^{\rm exp}_{p})$ and
$\Delta E_{\rm iso} \equiv (\log E^{\rm obs}_{\rm iso} - 
\log E^{\rm exp}_{\rm iso})$.
Here, $L^{\rm obs}_{p}$ ($E^{\rm obs}_{\rm iso}$) is the observed value, 
and $L_p^{\rm exp}$ ($E^{\rm exp}_{\rm iso}$) is the expected quantity
from Yonetoku and Amati relations for observed $E_p$, respectively.
We can see that there is no correlation between $\Delta L_{p}$ 
and $\Delta E_{iso}$.
We performed a linear partial correlation test for these 31~samples,
and found the partial correlation coefficient is quite small of 
$\rho_{L_{p}~E_{iso} \cdot E_p}=0.379$. 
Here $\rho_{12 \cdot 3}$ means the correlation coefficient between 
the first and the second parameters after fixing the third parameter
which is controlling two correlations 
(the parameter 1--3 and 2--3 relations). 
This fact indicates that two distance indicators may be independent
from each other. In the next section (\S-3), 
we investigate whether the cosmological parameters measured with 
two independent relations are consistent or not.
After that, in \S-4, we explore a hidden parameter 
which is the cause of the intrinsic dispersion of 
Amati and Yonetoku relations.

\section{Constraints on Cosmological Parameters}

We apply these Amati and Yonetoku relations to 29 GRBs with high
redshifts, $1.9 < z < 5.6$, to determine the luminosity distance
as a function of $z$. 
 We include not only measurement errors of $E_{iso}$ ($L_p$), 
and $E_p$ but also systematic errors we estimated in \S-2 and the 1-sigma 
uncertainties in Eq.~(\ref{eq:Amati}) (Eq.~(\ref{eq:Yonetoku})) to 
estimate luminosity distance \citep{Schaefer2007}.
For example, in case of Amati relation, we define $\mu_0(z_i)$ and 
$\sigma_{\mu_0,z_i}$ as below,
\begin{eqnarray}
\mu_0(z_i)&=&\frac{5}{2}\left(A+B\log{E_p}-\log(\frac{4\pi S_{bol}}{1+z})\right),\\
\sigma^2_{\mu_0,z_i}&=&\frac{25}{4}\Big[\sigma_A^2
+(\sigma_B\log{E_p})^2+\sigma_{sys}^2 \\
& &\mbox{ }+(0.434B\sigma_{Ep}/{E_p})^2+(0.434\sigma_{S_{bol}}/S_{bol})^2\Big].\nonumber
\end{eqnarray}

\begin{figure}
\includegraphics[width=84mm]{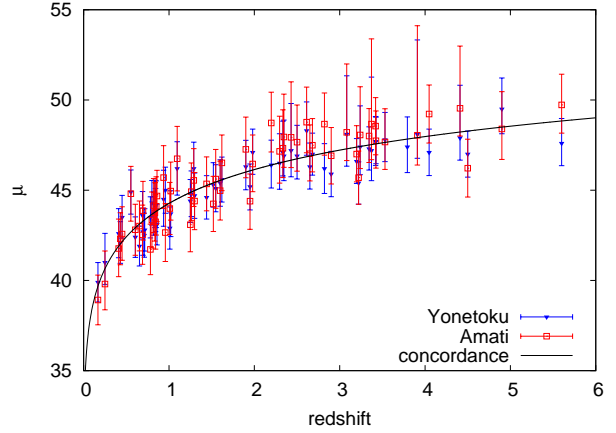}
\vspace{0pt}
\caption{Extended Hubble diagram from Yonetoku relation (blue) 
and Amati relation (red). A systematic difference seems to exist 
in high redshift, although it doesn't seem in low redshift GRBs.}
\label{fig3}
\end{figure}

Fig.~\ref{fig3} shows an extended Hubble
diagram up to $z = 5.6$ from Amati relation (red) and Yonetoku
relation (blue). A systematic difference between red and green points
seems to exist especially in high-redshift region.

Then we derive constraints on cosmological parameters.
In the $\Lambda$-CDM model with
$\Omega_k = \Omega_m + \Omega_{\Lambda} - 1$, 
the luminosity distance is given by,
\begin{eqnarray}
&& {d_L^{\rm th}}(z, \Omega_m, \Omega_{\Lambda})\! \nonumber \\
&& ~~~ = \! (1+z)\left\{
\begin{array}{ll}
\frac{c}{H_0 \sqrt{\Omega_k}} \sin(\sqrt{\Omega_k}F(z))
& \mbox{if}~~\Omega_k > 0\\
\frac{c}{H_0 \sqrt{-\Omega_k}} \sinh(\sqrt{-\Omega_k}F(z))
& \mbox{if}~~\Omega_k < 0\\
\frac{c}{H_0} F(z)
& \mbox{if}~~ \Omega_k = 0\\
\end{array}
\right.
\end{eqnarray}
with
\begin{equation}
F(z)
= \int_0^z dz' \,
  \Big[\Omega_m (1+z')^3 - \Omega_k (1+z')^2 + \Omega_{\Lambda}
  \Big]^{-1/2}.
\label{F(z)}
\end{equation}
The likelihood contour is defined by,
\begin{equation}
\Delta \chi^2
= \sum_{i}
  \left\{
  \frac{\mu_0({z_i}) - \mu^{\rm th}(z_i, \Omega_m, \Omega_{\Lambda},)}
       {\sigma_{\mu_{0, z_i}}}
  \right\}^2
  - \chi_{\rm best}^2,
\label{eq:contour1}
\end{equation}
where
$\mu^{\rm th}(z_i, \Omega_m, \Omega_{\Lambda})
= 5 \log( d_L^{\rm th}/{\rm Mpc}) + 25$
and $\chi_{\rm best}^{2}$ represents the chi-square value for
the best-fit parameter set of $\Omega_m$ and $\Omega_{\Lambda}$.

\begin{figure}
\vspace{0pt}
\includegraphics[width=70mm]{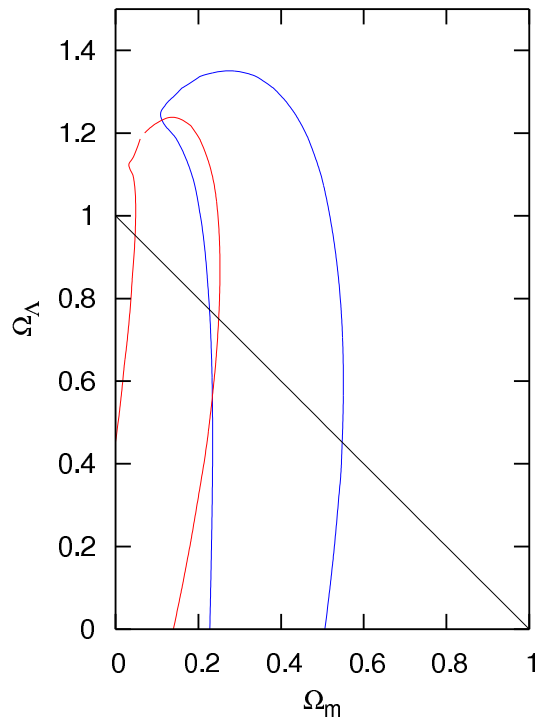}
\caption{Constraints on ($\Omega_m,\Omega_{\Lambda}$) plane from 
Amati relation (red) and Yonetoku relation (blue). The contours correspond
to 68.3\% confidence regions, and
black solid line represents the flat universe.
They are slightly different, although they are consistent in 
2-$\sigma$ level. See also Table~\ref{tab1}.}
\label{fig4}
\end{figure}

In Fig.~\ref{fig4}, we show the likelihood contour from
Amati (red) and Yonetoku relations (blue), and the best-fit
values with 1-$\sigma$ errors are shown in Table~\ref{tab1}.
Interestingly, they are slightly different, although they 
are consistent in 2-$\sigma$ level.

\begin{table}
\begin{center}
\begin{tabular}[htb]{|c|c|c|c|}
\hline
& $\Omega_m$ & $\Omega_{\Lambda}$ & $\chi_{\nu}^2$  \\
\hline
Amati & $0.10^{+0.15}_{--}$ & $1.15^{0.09}_{--}$ & 11.46/27 \\
Amati (flat) & $0.12^{+0.13}_{-0.07}$ & - & 11.69/28 \\
\hline
Yonetoku & $0.25^{+0.31}_{-0.15}$ & $1.24^{+0.11}_{--}$ & 9.36/28 \\
Yonetoku (flat) & $0.34^{+0.19}_{-0.14}$ & - & 9.71/29 \\ 
\hline
$E_p$-$T_L$-$L_p$ & $0.17^{+0.15}_{-0.08}$ & $1.21^{+0.07}_{-0.61}$ & 16.59/27 \\
$E_p$-$T_L$-$L_p$ (flat) & $0.24^{+0.11}_{-0.09}$ & - & 17.50/28 \\
\hline
\end{tabular}
\caption{Constraints on ($\Omega_m,\Omega_{\Lambda}$) in non-flat
and flat universe with 1-$\sigma$ errors from Amati, Yonetoku and
$E_p$-$T_L$-$L_p$ relations and their reduced chi squares. Constraints from Amati and Yonetoku
relations are inconsistent in 1-$\sigma$ level.}
\label{tab1}
\end{center}
\end{table}

\section{New relation}

In this section, we seek for a hidden parameter which reduces
the systematic error in the relation.  
Note that the discussion in this section does not concern with 
independence of the two relations.
If we could find this parameter, we can make strong constraint on cosmological 
parameters, because the systematic errors in the Amati and Yonetoku relations are larger than 
their measurement errors. \citep{Ghirlanda2006,Firmani2006b}
There are some studies using jet break time in afterglow as such a
parameter \citep{Ghirlanda2004, Liang2005}, but the number of GRBs which 
have observed jet is small. Here we restrict our discussion in a prompt emission property.
In the past studies, prompt emission was characterized by a time scale,
the duration of most intense parts of the GRB ($T_{0.45}$)
\citep{Firmani2006b, Rossi2008, Collazzi2008}.  Here we adopt a time scale called
luminosity time $T_L$ introduced by \citet{Willingale2007} as,
\begin{equation}
T_L = \frac{E_{\rm iso}}{L_p} = \frac{S_{\rm bol}}{(1+z) F_p}.
\end{equation}
The luminosity time does not depend on detector's energy band because it 
is defined by bolometric flux and fluence.

\begin{figure}
\includegraphics[height=84mm,angle=270]{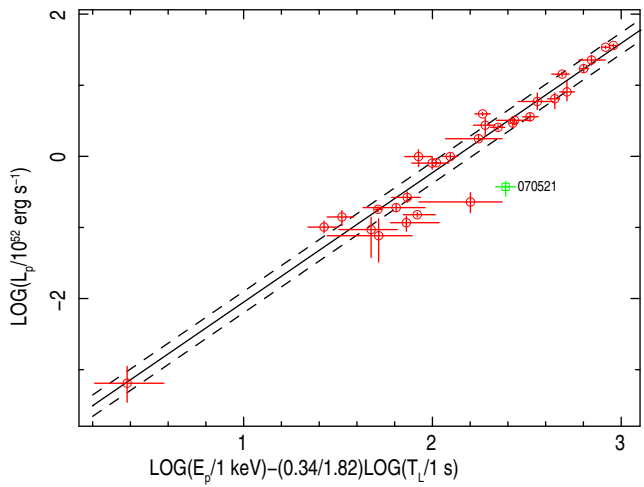}
\vspace{0pt}
\caption{The $E_p / T_L^{0.187}$ and $L_p$ in 31 GRBs with
$z < 1.62$. The correlation is improved than Amati and Yonetoku
relations. The solid line shows the best-fit curve without
one outlier (green square:GRB070521), and dashed lines represent
the 1-sigma region. The Pearson correlation coefficient is 0.971 and
reduced chi-square is $\chi^2 = 26.9/27$ with the systematic error 
$\sigma_{sys} = 0.15$ which is comparable
to the measurement error.}
\label{fig5}
\end{figure}

We assume the correlation among $E_p$, $T_L$ and $L_p$ to be of
the form, $\log{L_p} \equiv A + B \log{E_p} + C \log{T_L}$.
Then we obtain, 
\begin{equation}
\frac{L_p}{10^{52}~\rm{erg~s^{-1}}}
= 10^{-3.87 \pm 0.19}
  \left(\frac{E_p}{1~\rm{keV}}\right)^{1.82 \pm 0.08}
  \left(\frac{T_L}{1~\rm{s}}\right)^{-0.34 \pm 0.09},
\label{eq:new}
\end{equation}
from low redshift 30 GRBs in $0.16 < z < 1.7$. In Fig.~\ref{fig5}
we show $E_p$-$T_L$-$L_p$ relation. The Pearson correlation coefficient
is 0.971, the reduced chi-square is $\chi^2 = 26.9/27$ with
the systematic error $\sigma^{sys} = 0.15$. 
We include not only errors in $L_{p}$ but also errors in 
$E_p$, and $T_L$ so the chi-square function is defined as
$
\chi^2(A,B,C)=\Sigma({\log{L_{p}^{obs}}-A-B\log{E_p}-C\log{T_L}})^2/(\sigma_{meas}^2+\sigma_{sys}^2)
$
where the weighting factor 
$\sigma_{meas}^2=(1+2C)\sigma^2_{\log{L_{p}}}+(B\sigma_{\log{E_p}})^2+(C\sigma_{\log{T_L}})^2$ 
, and $\sigma_{\log{TL}}$ is estimated by using the error propagation 
equation without a crossterm between $L_p$ and $E_{iso}$. The factor 2C in the front of 
$\sigma^2_{\log{L_p}}$ comes from the fact that the definition of $T_L$ 
includes $L_p$. Note that the contribution of the additional error term 
$\sigma_{T_L}$ to chi-square value is little because of its small slope.
Thus, we conclude that the additional 
term really improves relations.
The systematic error is substantially reduced compared to those
of Amati and Yonetoku relations, and now comparable to
the measurement error. Thus this relation could be regarded
as "Fundamental plane" \citep{Djorgovski1983} of GRB prompt emission.

Here we excluded one outlier, GRB070521, from the fitting
of Eq.~(\ref{eq:new}). Actually, the host galaxy of GRB070521
is detected inside an error circle of XRT by
\citet{Hattori07_GCN6444} using Subaru Telescope after 40 minutes
from the trigger, but they couldn't detect bright afterglow.
Thus, the real redshift may be larger, which is why we exclude
GRB070521 from our analysis. 
Note that the probability of the miss identification of the host galaxy from only 
XRT observation is about 7\%\citep{Cobb2008}. Futhermore there might be 
 another population of GRBs so that  the secure classification of GRBs is one of 
 the current issue of GRB's study.
For more detailed discussions about
$E_p$-$T_L$-$L_p$ relation, see \citet{Tsutsui2008d}.

We show a possible derivation of the new relation (Eq.(\ref{eq:new})
under the photospheric model of the prompt emission of GRBs
( \citet{Ioka2007} and references therein).
The luminosity is given by
\begin{equation}
 L\propto r^2\Gamma^2{T'}^4
\label{eq:photoshere}
\end{equation}
where $r$, $\Gamma$ and $T'$ are the photospheric radius, the gamma
factor and co-moving temperature of the photosphere, respectively.
Since \citet{Ioka2007} assume that the energy is supplied by the relativistic collision of
the rapid shell of mass $m_r$ and Lorenz factor $\gamma_r$ with the slow shell of
$m_s$ and $\gamma_s$. Then under the perfectly inelastic collision model,
 $\Gamma$ is given by
$
 \Gamma^2=(m_r\gamma_r+m_s\gamma_s)/(m_r/\gamma_r+m_s/\gamma_s)
$.
Since $\gamma_r\gg\gamma_s$ and $m_r\gamma_r+m_s\gamma_s\propto E_{iso}$,
we can reduce $\Gamma^2\propto E_{iso}\gamma_s/m_s$.
If we regard $E_p\sim\Gamma T'$, we can rewrite Eq.(\ref{eq:photoshere}) as
$
 L\propto r^2E_p^4/\Gamma^2\propto r^2m_sE_p^4/E_{iso}\gamma_s.
$
Now let us assume $r$, $m_s$ and $\gamma_s^2T_L$ are constants.
Then we have
$
 L\propto E_p^2T_L^{-0.25}.
$
The above relation is essentially the same as  Eq.(\ref{eq:new}) if we
consider 2-$\sigma$ error of the power index in Eq.(\ref{eq:new}).

Possible reasons for the above three assumptions are as follows.
In \citet{Ioka2007} model $r$ is similar to the radius of the progenitor
star so that it could be constant. $T_L$ can be regarded as the effective 
duration of the burst. Then  $c\gamma_s^2T_L$ is the radius that the last rapid
shell catches the slow shell and we expect that this is also the order of
the radius of the progenitor star, which is constant. We have no reason why $m_s$
is constant. However if $m_s$ obeys the log normal distribution like the
other observables in GRBs we may regard it essentially constant.
If these assumptions are reasonable, the new relation
(Eq.(\ref{eq:new})) could be derived in the photospheric model of the prompt emission of 
GRBs like \citet{Ioka2007}.

Finally we use this new relation to put constraints
on cosmological parameters. 
Here again we include not only measurement errors of $L_p$, 
$E_p$, $T_L$ but also systematic errors and the 1-sigma 
uncertainties in Eq.(\ref{eq:new}) to estimate luminosity distance \citep{Schaefer2007}.

The concordance cosmology is still consistent
in 1-$\sigma$ level. The constraints on cosmological parameters are 
$(\Omega_m, \Omega_{\Lambda}) =
 (0.17^{+0.15}_{-0.08}, 1.21^{+0.07}_{-0.61})$ and
$\chi_{\nu}^2 = 16.59/27$ for non-flat universe,
$\Omega_m=0.24^{+0.11}_{-0.09}$ and $\chi_{\nu}^2 = 17.50/28$
for flat universe. (See Table~\ref{tab1}.)
\begin{figure}
\includegraphics[width=84mm]{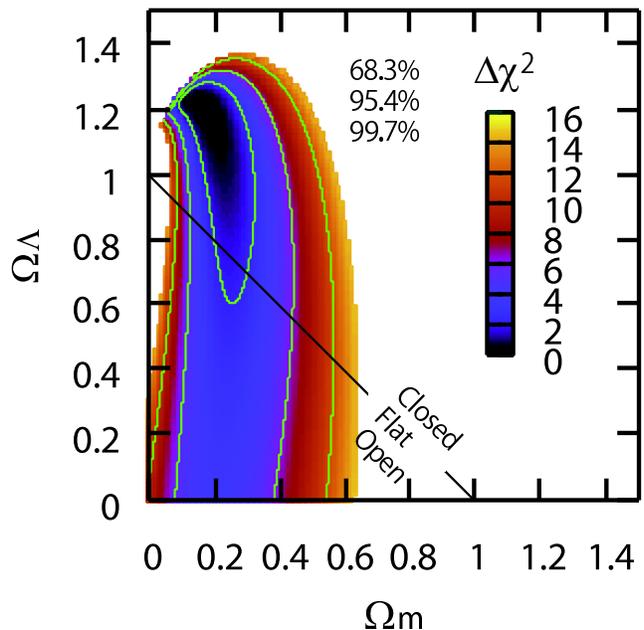}
\vspace{0pt}
\caption{Constraint on ($\Omega_m,\Omega_{\Lambda}$) plane from 
$E_p$-$T_L$-$L_p$ relation. See also Table~\ref{tab1}.}
\label{fig6}
\end{figure}

\section{Discussion}
Recently some authors extended Hubble diagram up to $z \sim 6$
using various luminosity indicators
\citep{Amati2008,Liang2008,Schaefer2007,Firmani2006a,Ghirlanda2006}.
\citet{Ghirlanda2006}, \citet{Firmani2006a}, and \citet{Schaefer2007} are pioneering works for 
GRB cosmology, but they are caught in circularity problem because there 
are few GRBs at low redshift.

\citet{Schaefer2007} obtained cosmological constraints from 
lag ($\tau_{\rm{lag}}$)-luminosity relation,
variability ($V$)-luminosity relation,
$E_p$-jet collimated energy ($E_{\gamma}$) relation (so called Ghirlanda 
relation),
minimum rise time ($\tau_{\rm{RT}}$)-luminosity relation,
and Yonetoku relation. \citet{Liang2008} calibrated these
relations by luminosity distances from SNIa observations. 

However, \citet{Tsutsui2008a} showed redshift dependence of
$\tau_{\rm{lag}}$-$L_p$ relation analyzing 565 BASTE GRB samples
with pseudo-redshifts estimated by Yonetoku relation. This suggests that
 the relation cannot be used as a distance indicator.
The dependence might come from the fact that $\tau_{\rm lag}$s
were evaluated from several fixed energy bands depending on the detectors
and they are different energy in GRB rest frame ,so it suffer from K-correction problem. 
This argument applies to $V$s and $\tau_{\rm{RT}}$s.

Although $E_p$-$E_{\gamma}$ relation and Liang \& Zhang relation have 
much smaller systematic errors than both Yonetoku and Amati relations,
there are many GRBs without a jet break or
with multiple jet breaks (missing or multiple jet break problem)
so that it is not certain whether the jet break can be used
to characterize GRB emission.
In contrast, our new relation does not suffer from these problems,
because it is totally defined by the prompt emission property and it has 
as a small systematic error as $E_p$-$E_{\gamma}$ and Liang \& Zhang relations.
Thus we could expect that the new relation would put GRB cosmology
to the next promising stage, as Phillips relation \citep{Phillips1993} and Fundamental
plane \citep{Djorgovski1983} did for SNeIa and elliptical galaxies. 
However, we must
emphasize that we need detailed studies of the new relation with
much larger number of GRBs and examination of systematic errors
in order for GRB to be regarded as a reliable tool for cosmology
like SNIa, cosmic microwave background, baryon acoustic oscillation
and gravitational lens. Now ongoing Missions like \swift, \fermi and 
\suzaku, and the collaboration of many observer on ground will promise the progression of GRB cosmology.

\section*{Acknowledgments}

This work is supported in part by the Grant-in-Aid from the 
Ministry of Education, Culture, Sports, Science and Technology
(MEXT) of Japan,  No.19540283, No.19047004(TN),
and  No.20674002 (DY) and  by the Grant-in-Aid for the global
COE program
{\it The Next Generation of Physics, Spun from Universality and Emergence}
from MEXT of Japan. KT is supported in part by a Grant-in-Aid from the Japan Society for the Promotion of Science (JSPS) Fellows and the global COE programs, "Quest for Fundamental Principles in the Universe: from Particles to the Solar System and the Cosmos" at Nagoya University.

\end{document}